\documentclass[12pt]{iopart}
\usepackage{graphicx}
\usepackage{iopams}
\usepackage{caption}

\usepackage[bookmarks=false]{hyperref}
\hypersetup{colorlinks=true,citecolor=blue,linkcolor=red,%
urlcolor=blue,pdfstartview=FitH,bookmarksopen=true}

\newcommand{\openone}{\mathbb{I}}
\newcommand{\ket}[1]{\mbox{$ | #1 \rangle $}}

\newcommand{\cM}{\mathcal{M}}

\newcommand{\cH}{\mathbb{H}}
\newcommand{\cR}{\mathbb{R}}

\newcommand{\text}[1]{\mathrm{#1}}
\newcommand{\by}{{\mathbf y}}
\newcommand{\bv}{{\mathbf v}}
\newcommand{\bz}{{\mathbf z}}
\newcommand{\bx}{{\mathbf x}}

\newcommand{\minover}[1]{\min_{#1}}
\newcommand{\subto}{\text{s.t.}}

\begin{document}

\title{Corrupted sensing quantum state tomography}

\author{Mengru Ma}
\address{Key Laboratory of Advanced Optoelectronic Quantum Architecture and Measurement (MOE), School of Physics, Beijing Institute of Technology, Beijing 100081, China}

\author{Jiangwei Shang}
\address{Key Laboratory of Advanced Optoelectronic Quantum Architecture and Measurement (MOE), School of Physics, Beijing Institute of Technology, Beijing 100081, China}
\ead{jiangwei.shang@bit.edu.cn}
\vspace{10pt}
\date{\today}
%

\begin{abstract}
The reliable characterization of quantum states as well as any potential noise in various quantum systems is crucial for advancing quantum technologies. In this work we propose the concept of corrupted sensing quantum state tomography which enables the simultaneous reconstruction of quantum states and structured noise with the aid of simple Pauli measurements only. Without additional prior information, we investigate the reliability and robustness of the framework. The power of our protocol is demonstrated by assuming sparse Gaussian and Poisson noise for low-rank state tomography. In particular, our approach is able to achieve a high quality of the recovery with incomplete sets of measurements and is also suitable for performance improvement of large quantum systems. It is envisaged that the techniques can become a practical tool to greatly reduce the cost and computational effort for quantum tomography in  noisy quantum systems.
\end{abstract}
\vspace{2pc}

\section{Introduction}
The realization of quantum information processing depends on the precise characterization of the quantum systems~\cite{eisert2020quantum}. Quantum state tomography (QST) is a well-established approach to reconstruct a general quantum state (either pure or mixed) through a series of  measurements performed on identically prepared input states~\cite{RevModPhys.29.74,PhysRevA.56.1788,cramer2010efficient,PhysRevA.85.052120,Shang.etal2017,torlai2018neural}. However, standard tomography is extremely resource intensive as the number of measurement settings required scales exponentially with the size of the system. There has been an increasing effort to develop techniques that minimize the resource necessary  for tomography. To this end, the methodology of compressed sensing has been applied to the problem of quantum tomography. In the pioneering work of Refs.~\cite{Gross.etal2010, Gross2011}, it was proved that fairly pure quantum states, described with an unknown density matrix of dimension $d$ and rank $r$, can be reconstructed using $O(rd\log^2d)$ measurement settings only, while standard methods including QST require at least $d^2$ settings.

Meanwhile, noise is almost inevitable in all quantum information processing tasks and  it is one of the main obstacles toward realizing universal quantum computing~\cite{Preskill2018quantum, science.abb2823}. As quantum computers tend to approach the fault-tolerant regime, especially as the overhead of full error correction and fault tolerance is beyond the capability of current hardware, noise diagnosis and characterization then become increasingly important, yet unfortunately intractable~\cite{PhysRevX.9.041053, PhysRevLett.124.130501}. Although techniques such as dynamical decoupling~\cite{Viola1999}, Pauli frame randomization~\cite{Knill2005, Ware2021}, and randomized compiling~\cite{Emerson2016, Gu2023, chen2023learnability} can be used to transform a general quantum channel into a Pauli channel, the estimation of noise is still inefficient. Therefore, structural assumptions about noise is imperative to leverage the burden in noise description. For instance, $n$-qubit Pauli channel with bounded degree correlations can be learned efficiently in time ${poly}(n)$~\cite{Flammia2020,Harper2020}, and Pauli channels with at most $s$ nonzero Pauli error rate ($s$-sparse) are also discussed in efficient algorithms with the classical processing time being $O(sn^2)$ \cite{Harper2021}.

Not merely noise channels, noise may appear in many different forms. For instance, measurements in quantum mechanics are inherently probabilistic, leading to statistical noise in any measurement with a finite number of input states~\cite{langford2013errors}. Besides, when preparing the initial states and performing measurements, the difference between the ideal and actual probabilities results in state preparation and measurement (SPAM) noise, which is the main source of systematic noise~\cite{palmieri2020experimental, PhysRevResearch.3.033285}.
A common idea to deal with noise is to quantify the effect of noise on experimental outcomes, so as to realize the regulation and control of the system. Moreover, to recover the noise \emph{simultaneously} with the state is preferable and indeed possible if both the state and noise own certain structures.

Very recently, the problem of simultaneous tomography has garnered much attention. Based on the discussion of gauge fixing with prior information, an algorithm for simultaneous reconstruction of quantum state and SPAM noise was presented in~\cite{jayakumar2023universal}. It is said that conditions for ensuring simultaneous tomography include requiring the output of noise matrix to correlate with the input, preventing the quantum state from being maximally mixed, and making the set of measurement operators linearly independent.

In signal processing, the technique of \emph{corrupted sensing}~\cite{li2013compressed, Nguyen6376185, foygel2014corrupted, mccoy2014sharp, Chen.Liu2017, Chen.Liu2019} concerns the problem of recovering a structured signal from a relatively small number of noisy (corrupted) measurements. Since this problem is generally ill-posed, tractable recovery is only possible when both the signal and corruption are suitably structured.
Motivated by the concept of corrupted sensing, in this work we present a general framework to reconstruct the quantum state as well as noise simultaneously. Our protocol not only applies to scenarios where the data obtained by measurements may be corrupted (or there is a deviation between the obtained data and the true expected values), but also provides a way to characterize certain noise. We shall employ experimentally friendly Pauli measurements~\cite{liu2011universal, Flammia2011} and characterize the state and noise with suitable norm functions.

The rest of the paper is organized as follows. We first introduce notations and background information in section~\ref{Sec:CSPm}, then an intuitive overview of the recovery method is presented in section~\ref{Sec:algorithm}. In section~\ref{Sec:Applications}, various applications are shown together with the details of the validation results. Finally, we provide some discussions in
section~\ref{Sec:discussion} and conclude with an outlook in section~\ref{Sec:Summary}.

\section{Corrupted sensing QST with Pauli measurements}\label{Sec:CSPm}
Considering an $n$-qubit quantum system with dimension ${d=2^n}$, the unknown state of the system, denoted by $\rho$, is a rank-$r$ ($r\!\leq\!d$) density matrix that satisfies ${\Tr(\rho)=1}$ and ${\rho\succeq 0}$.
An $n$-qubit Pauli operator takes on the general form
\begin{equation}
  P=\bigotimes_{i=1}^n\sigma_i\,,
\end{equation}
where ${\sigma_i\in\{\openone, \sigma_x, \sigma_y, \sigma_z\}}$. Here, $\sigma_x, \sigma_y, \sigma_z$ are the three Pauli matrices, and $\openone$ represents the identity matrix. In total there are ${d^2=4^n}$ such Pauli operators, which form a set denoted by $\mathcal{P}$.

In general, the simultaneous reconstruction of quantum state and structured noise consists of the following two steps: First select Pauli operators at random to measure the quantum state and obtain the noisy data; then choose a suitable convex optimization algorithm for data post-processing to get the estimations of the state and noise.

To be specific, the scheme of corrupted sensing QST proceeds as follows.
Choose $m$ Pauli operators $\{P_1,P_2,\cdots,P_m\}$ uniformly at random from $\mathcal{P}$, and for each ${k\in \{1,\cdots,m\}}$, measure the expectation values $\Tr(P_k\rho)$. These Pauli operators are chosen without replacement, and once selected they remain fixed.
To get an estimate of the expectation value $\Tr(P_k\rho)$, we use $N$ copies of the state $\rho$.

Let $\cH^d$ denote the set of ${d\times d}$ Hermitian matrices, and define the linear map ${\cM:\cH^d\to \cR^m}$ for all $P_k$s as
\begin{equation}\label{eq:getdata}
  [\cM(\rho)]_k=\Tr(P_k\rho)\,.
\end{equation}
Then, the output of the entire measurement process can be written as a vector
\begin{equation}\label{eq:corSens}
  \by=\cM(\rho)+\bv+\bz\,.
\end{equation}
Here the structured noise (or structured corruption) is modeled as a stochastic vector $\bv\in\cR^m$, which is a general consideration as noise can manifest in any process. And $\bz\in\cR^m$ is any other kind of unstructured noise including statistical noise. In the context, \emph{structured noise} refers to the noise with a defined pattern that can be modeled using an appropriate norm, such as the sparse noise, which exhibits low complexity when measured by the $\ell_1$-norm. In contrast, \emph{unstructured noise} lacks a specific pattern, making it difficult to model or control directly.
In particular, if there's no corruption, i.e., ${\bv=0}$, the model in~(\ref{eq:corSens}) reduces
to the standard compressed sensing problem~\cite{Gross.etal2010, Gross2011, Flammia.etal2012}.

\begin{figure}[t]
\centering
    \includegraphics[width=.5\columnwidth]{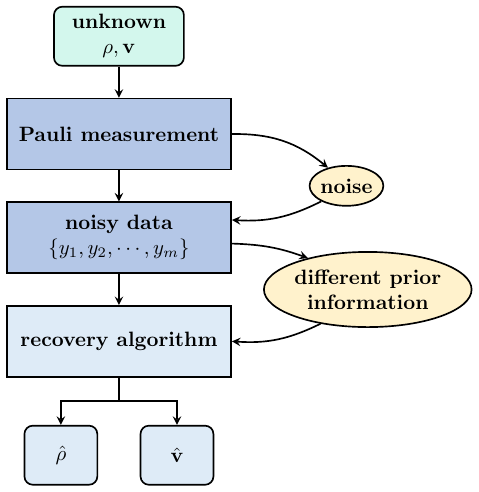}
    \caption{Schematic procedure of the corrupted sensing quantum state tomography. For the unknown state $\rho$ and noise $\bv$, Pauli measurements are employed to get the noisy data ${\by=\{y_1,y_2,\cdots,y_m\}}$. With different prior information, one can choose various recovery algorithms to get the reconstructed state $\hat{\rho}$ and noise $\hat{\bv}$; see~(\ref{eq:cvxgeneral}), ~(\ref{eq:Ala1}), and~(\ref{eq:Alb1}).}
    \label{fig1}
\end{figure}

Generally speaking, the problem in~(\ref{eq:corSens}) is ill-posed, and tractable recovery is only possible when both the state $\rho$ and the noise $\bv$ are suitably structured.
See figure~\ref{fig1} for a schematic framework of the corrupted sensing quantum state tomography. By randomly selecting $m$ Pauli operators to measure the quantum state, an estimation of both the state and noise from the acquired noisy data is then performed. Here we consider the general setting where no prior information about the quantum state $\rho$ or the structured noise $\bv$ is taken into account. Two other possible settings with different prior information are discussed in~\ref{App:Constrained} and~\ref{App:Penalized} respectively.

%

\section{The corrupted sensing QST estimator}\label{Sec:algorithm}
We consider the general case where the structure of the state and noise can be characterized by suitable norm functions.
Typical examples of such structures include low-rank  matrices and sparse vectors. The definition of sparseness indicates that in a sparse matrix or vector, most of the elements are zero. However, be noted that this is not sufficient in all cases, and it leaves open the question of how sparseness should be measured in general.
Hereafter, let $f(\cdot)$ and $g(\cdot)$ denote the suitable norms which fully characterize the structures of the state and noise respectively.

The reconstruction of the unknown state $\rho$ and structured noise $\bv$ without prior assumptions can be formulated as the following optimization problem
\begin{equation}\label{eq:cvxgeneral}
  \minover{\tilde{\rho},\tilde{\bv}}\frac{1}{2}\|\by-\cM(\tilde{\rho})-\tilde{\bv}\|_2^2+\tau_1\cdot f(\tilde{\rho})+\tau_2\cdot g(\tilde{\bv})\,,
\end{equation}
where ${\tau_1,\tau_2>0}$ are regularization parameters, and $\tilde{\rho}$ and $\tilde{\bv}$ represent the variables to be solved. Let ${\kappa \geq 1}$ be a real number, and the $\ell_{\kappa}$-norm of vector $\mathbf{x}=\left(x_1, \ldots, x_{\alpha}\right)\in\cR^{\alpha}$ is
\begin{equation}
\|\bx\|_{\kappa}:=\left(\sum_{i=1}^{\alpha}\left|x_i\right|^{\kappa}\right)^{1 / {\kappa}}\,.
\end{equation}
For ${\kappa=1}$, we get the $\ell_1$-norm; for ${\kappa=2}$, it is the $\ell_2$-norm. The intuition of the problem is to find $\tilde{\rho}, \tilde{\bv}$  which fit the data $\by$ while minimizing the least-squares linear regression with suitable norm regularizations.

Here we consider minimizing the trace norm $\|X\|_{\tr}=\Tr(\sqrt{X^{\dagger} X})$, which serves as a convex surrogate for minimizing the rank of $X$ in quantum state estimation. This choice reflects the assumption that the quantum state under consideration is of low-rank. On the other hand, the $\ell_1$-norm is chosen for the sparse noise $\bv$, as regularization with the $\ell_1$-norm can be used as a heuristic for finding sparse
solutions~\cite{boyd2004convex,cpa.20124,nielsen2015neural}. Moreover, the $\ell_1$-norm is the best convex approximation of the non-convex $\ell_0$-norm for sparse recovery, where the $\ell_0$-norm, denoted by $\|\bx\|_0$, is the number of non-zero components of the
vector $\bx$~\cite{cpa.20132}. Therefore, the estimators $\hat{\rho}, \hat{\bv}$ are obtained by
\begin{equation}\label{eq:normgiven}
    (\hat{\rho},\hat{\bv}) = \mathop{\arg\min}_{\tilde{\rho} \succeq 0, \tilde{\bv}}  \frac{1}{2}\|\by - \cM(\tilde{\rho}) - \tilde{\bv}\|_2^2 + \tau_1 \cdot \|\tilde{\rho}\|_{\tr}+\tau_2 \cdot \|\tilde{\bv}\|_1, \  \tau_1, \tau_2 > 0\,,
\end{equation}
in which the residual ${\by-\cM(\tilde{\rho})-\tilde{\bv}}$ is measured by the $\ell_2$-norm\footnote{The $\ell_2$-norm is commonly used in various applications, such as the least-squares problems~\cite{boyd2004convex} or as a regularization term in machine learning to prevent overfitting~\cite{nielsen2015neural}.} and $\arg\min$ refers to the input variables at which a function is minimized~\cite{boyd2004convex}.
Whenever the trace of the resulting estimate of the quantum state is not equal to $1$, we renormalize it as ${\hat{\rho}/\Tr(\hat{\rho})\mapsto\hat{\rho}}$.
To quantify the goodness of the reconstruction, we employ the (squared) fidelity~\cite{Flammia2011}
\begin{equation}\label{Fid}
F(\rho,\hat{\rho})=\left(\Tr\sqrt{\sqrt{\hat{\rho}}\rho\sqrt{\hat{\rho}}}\right)^2\,,
\end{equation}
and the mean squared error (MSE)
\begin{equation}\label{MSE}
   T_{\text{MSE}} = \frac{1}{m}\sum_{i=1}^m (\bv_i - \hat{\bv}_i)^2\,
\end{equation}
for the estimators $\hat{\rho}$ and $\hat{\bv}$ respectively.

\section{Applications}\label{Sec:Applications}
Using Pauli measurements, we numerically simulate the reconstruction of ${n=5}$ qubit random states and $W$ states under the corruption of $s$-sparse statistical noise. Showcases for larger systems can be found in~\ref{App:7qubit}.
In light of the convex characteristic of the problem as in~(\ref{eq:normgiven}), we rely on the {\sc cvx} package~\cite{cvx,RevModPhys.96.045006} with
the solver \textsc{sdpt3}~\cite{SDPT3} for efficient numerical solutions. Random states are generated using the {\sc qetlab} package by first generating a pure state according to the Haar measure on a larger space and then tracing out the ancillary subspace~\cite{qetlab,RevModPhys.96.045006}. The code as well as the corresponding data are available on GitHub~\cite{code}.

\subsection{Corruption by sparse statistical noise}\label{subsec4.1}
\begin{figure*}[t]
\centering
    \includegraphics[width=.9\textwidth]{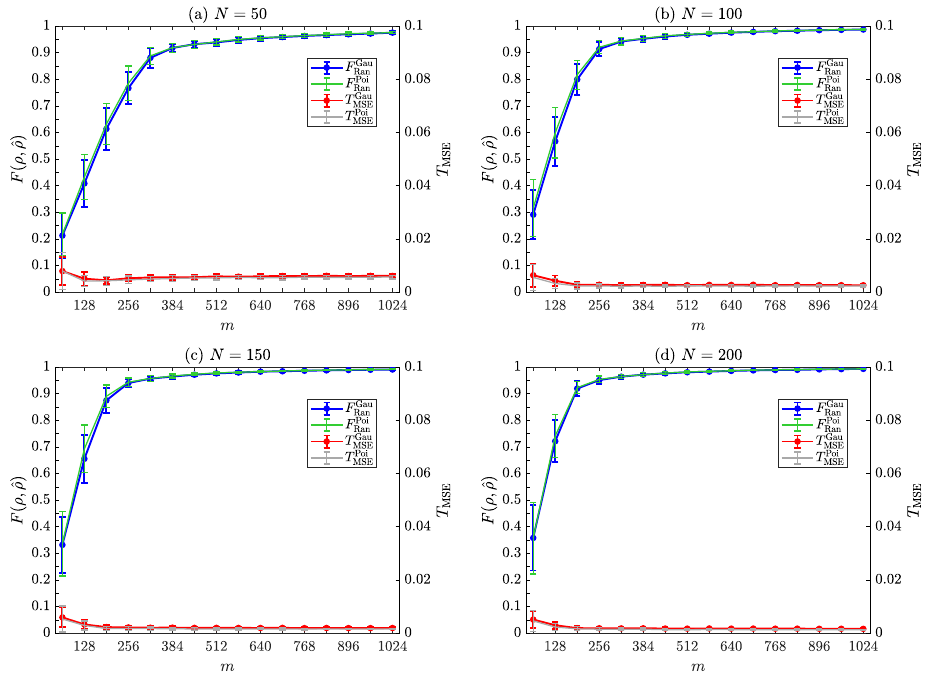}
    \caption{Fidelity $F(\rho,\hat{\rho})$ and MSE $T_{\text{MSE}}$ as functions of the number of sampled Pauli operators $m$ (ranging from $64$ to $1024$ with steps of $64$) over $120$ runs with ${n=5}$ qubits. Error bars are derived by considering $120$ different five-qubit random states, with $m$ Pauli operators randomly sampled for each state. The blue solid curve (green solid curve) represents the fidelity between the reconstructed state and true state in the case of sparse Gaussian (Poisson) noise. Meanwhile, the red solid curve (gray solid curve) represents the MSE between the reconstructed sparse Gaussian (Poisson) noise and true sparse Gaussian (Poisson) noise. The number of copies of the input random pure states used for each measurement in (a), (b), (c), and (d) are ${N=50, 100, 150}$, and $200$, respectively. Standard deviation of the sparse Gaussian noise and parameter of the sparse Poisson noise are both set to ${\sigma = \lambda = 1}$. Additionally, the regularization parameters are chosen as ${\tau_1=0.011m}$, ${\tau_2=0.16}$, and the sparsity level is taken as ${s=\lfloor 0.04m \rfloor}$.}
    \label{fig2}
\end{figure*}
\begin{figure}[t]
\centering
    \includegraphics[width=.5\columnwidth]{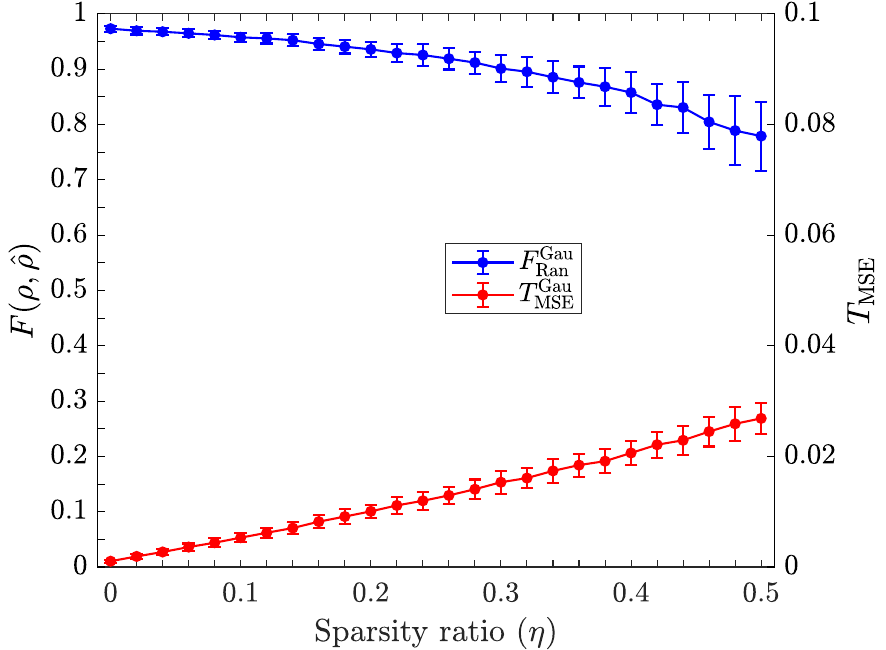}
    \caption{Fidelity $F(\rho,\hat{\rho})$ and MSE $T_{\text{MSE}}$ as functions of the sparsity ratio $\eta$ over $120$ runs with ${n=5}$ qubits. The number of copies of the input random pure states used for each measurement is specified as ${N=100}$, and the number of Pauli operators is fixed as ${m=512}$. Standard deviation of the sparse Gaussian noise is set to ${\sigma=1}$, and the regularization parameters are chosen as ${\tau_1=0.011m, \tau_2=0.16}$.}
    \label{fig_s}
\end{figure}

Due to the environment and measurement errors, Gaussian noise is unavoidable in any quantum information processing tasks.
Besides, in precision measurement and quantum optics, dark count rate is an important parameter of single-photon detectors which contributes as a main factor leading to error rates in tasks such as quantum communication~\cite{Stucki2009,PhysRevX.4.041041,RevModPhys.92.025002}. Dark count refers to the trigger of a detector in the absence of actual photons, which can usually be modeled as Poisson noise~\cite{ding2018pile,Li:19,10.1063/5.0226118,PhysRevA.78.055803}. If the detection efficiency of the photon in an experiment is high, and/or the dark count rate is small, then photon loss and dark counts can be considered as sparse Poisson noise.
In experiments, it is desirable to have a low dark count rate. Assuming a Gaussian distribution appropriate for the thermal noise, the expected dark count rate can be reduced to as low as $10^{-3}$ cps~\cite{miller2003demonstration}.

For the first application, we consider random pure states; see~\ref{App:Wstate} for the case of $W$ states. Figure~\ref{fig2} displays the fidelity $F(\rho,\hat{\rho})$ and the MSE $T_{\text{MSE}}$ as functions of the number of sampled Pauli operators $m$ (ranging from $64$ to $1024$ with steps of $64$) over $120$ runs. For each Pauli operator $P_k$, we take ${N = 50, 100, 150, 200}$ copies of the input random states in order to get the estimated value of $\Tr(P_k\rho)$. Sparse Gaussian noise $\bv$ is generated as corruption by constructing a vector with $s$ independent Gaussian entries of zero mean and standard deviation $\sigma$, while the remaining components of $\bv$ are set to zero. The generation of sparse Poisson noise is similar, with $\lambda$ as its parameter, which corresponds to the theoretical mean and variance of the Poisson entries. Several features are immediately available.

 Under sparse Gaussian noise, the fidelity $F(\rho,\hat{\rho})$ (blue solid curve) improves along with the increasing number of sampled Pauli operators $m$. For instance, in figure~\ref{fig2}(b), the fidelity achieves ${\sim 0.987}$ with ${m=1024}$ and ${N=100}$. However, a key result of our work is that fidelities as high as ${F(\rho,\hat{\rho})\approx 0.95}$ can be achieved with significantly fewer measurements, requiring only ${m\approx 37.5\%d^2}$ Pauli operators. In addition, a large number of samples prove advantageous in enhancing the precision and stability of the reconstruction. In figure~\ref{fig2}(a)-(d), achieving ${F(\rho,\hat{\rho})\approx 0.95}$ necessitates ${m\approx 640, 384, 320}$, and $256$, respectively. On top of that, the number of state copies $N$ has been increased to explore the convergence behavior of the algorithm, as detailed in~\ref{App:N}.

On the other hand, the MSEs $T_{\text{MSE}}$ between the reconstructed noise and true noise (red solid curve) are all in the order of $10^{-3}$ for figure~\ref{fig2}(a)-(d) as long as the fidelity of the corresponding reconstructed state reaches the threshold of $0.95$.
Expectedly, the MSE declines and stabilizes as $m$ and $N$ grow. Therefore, there exists a trade-off between the cost of measurement and the recovery accuracy.

Notably, sparse Poisson noise exhibits performance in reconstruction similar to those of sparse Gaussian noise under specific parameter settings, despite their different probability distributions and statistical characteristics. This reflects the universality of our reconstruction algorithm to statistical noise, providing further insights for selecting appropriate noise models.

Figure~\ref{fig_s} displays the fidelity and MSE as functions of the sparsity ratio. Here we consider random pure states with $s$-\emph{sparse} Gaussian noise, i.e., those with ${s=\lfloor \eta m \rfloor}$ nonzero elements. Correspondingly, the sparsity ratio $\eta$ quantifies the proportion of nonzero elements within the vector $\bv$. The change in $\eta$ leads to a variation in the sparsity of the noise, which then modifies the noisy data and affects the fidelity of the state. It is evident that with the increase of the sparsity ratio $\eta$, the fidelity decreases, accompanied by an escalation in MSE. For ${0\leq\eta\lesssim 0.14}$, the fidelity ${F(\rho,\hat{\rho})\gtrsim 0.95}$.

There are many factors affecting the setting of regularization parameters $\tau_1$ and $\tau_2$. And the reconstruction error relies on the choice of $\tau_1$ and $\tau_2$, as discussed in Theorem~1 of Ref.~\cite{Flammia.etal2012}, and Theorem~4 of Ref.~\cite{Chen.Liu2019}. In principle, the smaller $\tau_2$ is, the smaller the recovery error for $\bv$ will be. However, there is a trade-off between the recovery errors of the state $\rho$ and corrupted noise $\bv$. Drawing from these references and empirical observations, one can start by simply selecting several integer values (such as ${1,\ldots,10}$) to explore and refine the parameter range. Here, we adopt the parameter combination ${\tau_1=0.011m}$ and ${\tau_2=0.16}$, and leave the optimal parameter selection as an open problem.

\subsection{Robustness of the protocol}\label{subsec4.2}
\begin{figure}[t]
\centering
   \includegraphics[width=.5\columnwidth]{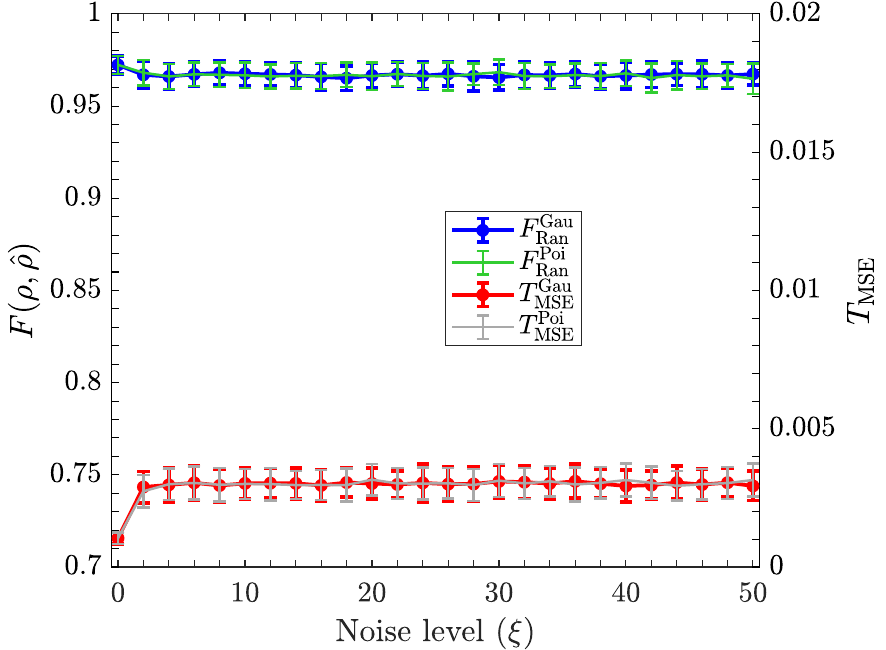}
    \caption{Fidelity $F(\rho,\hat{\rho})$ and MSE $T_{\text{MSE}}$ as functions of noise level $\xi$ over $120$ runs with ${n=5}$ qubits. For sparse Gaussian noise, ${\xi=\sigma}$, where $\sigma$ is the standard deviation. For sparse Poisson noise, ${\xi=\lambda}$, where $\lambda$ is the Poisson noise parameter. The number of copies of the input random pure states used for each measurement is specified as ${N=100}$ and the number of Pauli operators is fixed as ${m=512}$. The regularization parameters are chosen as ${\tau_1=0.011m, \tau_2=0.16}$. The sparsity level is taken as ${s=\lfloor 0.04m \rfloor}$.}
    \label{fig_NLevel}
\end{figure}
To demonstrate the algorithm's robustness, we show the variations in fidelity $F(\rho,\hat{\rho})$ and MSE $T_{\text{MSE}}$ with respect to the noise level of sparse Gaussian and Poisson noise in figure~\ref{fig_NLevel} by changing the corresponding parameters $\sigma$ and $\lambda$.
The results confirm that our protocol remains robust against different noise levels. Particularly, considering sparse Gaussian noise  with standard deviation ${\sigma=0}$ as corruption, it results in ${F(\rho,\hat{\rho})\approx0.97}$ and ${T_{\text{MSE}}\approx10^{-3}}$ due to systematic error. Distinct from corrupted noise, the systematic error may result from several factors, including inherent approximation errors due to the nature of the model or optimization strategy, biases introduced by the $\ell_1$ and $\ell_2$ regularization, and numerical errors in the calculation. As $\sigma$ grows, the fidelity consistently stays $\sim 0.96$, with ${T_{\text{MSE}}\approx 3\times 10^{-3}}$. The relative $\ell_2$-norm error of the noise reconstruction is displayed in~\ref{App:noiseError}. Consistent with the findings in figure~\ref{fig2}, the scenarios of the changes in sparse Poisson noise level exhibit similar impacts on fidelity and MSE to those of the sparse Gaussian noise.

\begin{figure*}[t]
\centering
    \includegraphics[width=.9\textwidth]{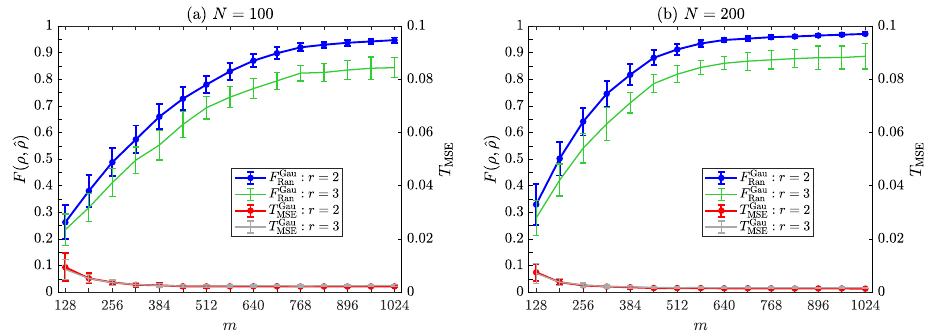}
    \caption{Fidelity $F(\rho,\hat{\rho})$ and MSE $T_{\text{MSE}}$ as functions of the number of sampled Pauli operators $m$ over $120$ runs of the input rank-$2$, rank-$3$ random states with $n=5$ qubits. The copy numbers used for each measurement are as follows: (a) $N=100$, and (b) $N=200$. Standard deviation of the sparse Gaussian noise is set to $\sigma=1$. And the regularization parameters are chosen as $\tau_1=0.0055m, \tau_2=0.15$. The sparsity level is taken as $s=\lfloor 0.04m \rfloor$.}
    \label{fig_r}
\end{figure*}

In reality, the input state $\rho$ may not be a pure state due to potential noise, resulting in higher-rank states. Using sparse Gaussian noise,
figure~\ref{fig_r} displays fidelity $F(\rho,\hat{\rho})$ and MSE $T_{\text{MSE}}$ as functions of the number of sampled Pauli operators $m$ for rank-$2$ and rank-$3$ input random states. For rank-$2$ random states, employing all available measurement operators yields fidelities reaching up to ${F(\rho, \hat{\rho})\approx 0.947}$ and ${F(\rho, \hat{\rho})\approx 0.97}$, with ${T_{\text{MSE}} \approx 2.3 \times 10^{-3}}$ and ${T_{\text{MSE}} \approx 1.5 \times 10^{-3}}$ in figure~\ref{fig_r}(a) and~(b), respectively. In comparison, for rank-$3$ random states, the fidelity ${F(\rho, \hat{\rho})}$ attains $0.887$, even with ${N=200}$ state copies and ${m=1024}$ measurement operators, accompanied by a corresponding MSE ${T_{\text{MSE}}\approx 1.7\times 10^{-3}}$. This suggests that our protocol performs more effectively in the case of lower-rank states. The regularization parameters $\tau_1,\tau_2$ are fine-tuned here, one can also achieve a higher fidelity by increasing the number of samples.

\begin{figure*}[t]
\centering
   \includegraphics[width=.9\textwidth]{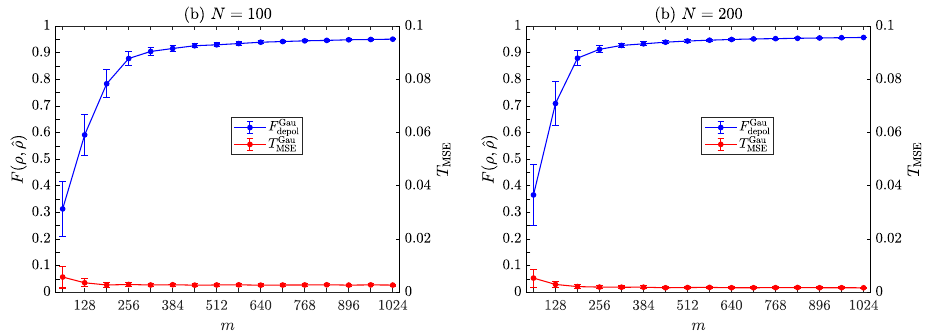}
\caption{Fidelity $F(\rho,\hat{\rho})$ and MSE $T_{\text{MSE}}$ as functions of the number of sampled Pauli operators $m$ over $120$ runs with $n=5$ qubits. The number of copies of the input random pure states with $1\%$ local depolarizing noise used for each measurement in (a), (b) are specified as ${N=100, 200}$ respectively. Standard deviation of the sparse Gaussian noise is set to ${\sigma=1}$. And the regularization parameters are chosen as ${\tau_1=0.011m, \tau_2=0.16}$. The sparsity level is taken as ${s=\lfloor 0.04m \rfloor}$.}
\label{fig_depol}
\end{figure*}

Furthermore, due to the existence of inevitable systematical noise, we shall consider noise to the state by applying independent and identical depolarizing channels to each of the $n$ qubits. The depolarizing channel
with strength ${\gamma\in[0,1]}$ acting on a single qubit $\rho$ is
\begin{equation}\label{depol}
  \mathcal{E}(\rho)=\frac{\gamma \openone}{2}+(1-\gamma)\rho\,.
\end{equation}
This means that, with probability $\gamma$ the qubit is replaced by a completely mixed state and otherwise it is left untouched. We assume very weak decoherence and set ${\gamma= 0.01}$.
For input random pure states with $1\%$ local depolarizing noise, figure~\ref{fig_depol} shows the fidelity ${F(\rho, \hat{\rho})}$ between the reconstructed state and the true state, as well as the MSE ${T_{\text{MSE}}}$ of sparse Gaussian noise, plotted as functions of the number of sampled Pauli operators $m$. The fidelity reaches ${F(\rho, \hat{\rho}) \approx 0.95}$ when ${m=960}$, as shown in figure~\ref{fig_depol}(a).  In contrast, with a reduced number of Pauli operators and $N=200$, the fidelity still surpasses $0.95$ at ${m=640}$, as shown in figure~\ref{fig_depol}(b),  with the corresponding MSE ${T_{\text{MSE}}\approx 1.7 \times 10^{-3}}$. Despite the presence of depolarizing noise, the protocol can still maintain effective performance with an incomplete set of measurement operators, thereby showcasing its robustness.

Finally, in~\ref{App:7qubit}, we show the protocol's potential for larger systems with an increasing number of qubits.

%

\section{Discussions}\label{Sec:discussion}
Although quantum tomography is a valuable tool in quantum information processing, it demands a substantial amount of resources. Therefore, various more efficient characterization methods have been developed, including for instance, our scheme, compressed sensing tomography~\cite{Gross.etal2010}, and the adaptive statistical framework based on Bayesian inference and Shannon's information~\cite{PhysRevA.85.052120}. Additionally, the Bayesian statistical model can be tailored to a variety of quantum systems with restricted measurement capabilities~\cite{lu2022bayesian}, and techniques relying on machine learning can be used to perform QST of highly entangled states with more than one hundred qubits~\cite{ torlai2018neural}.

An alternative approach forgoes full tomography of the state, with the classical shadow scheme~\cite{huang2020predicting} based on randomized measurements~\cite{elben2023randomized,PhysRevLett.131.090201} as a typical example. Although this method excels in predicting many properties of a quantum system, the classical shadow of $\rho$ may be unphysical as the inverted channel may not be physical. Moreover, a tomographically complete set of measurements is required to ensure that the quantum channel is invertible.

Beyond that, in settings where measurements are noisy or incomplete, QST based on convolutional neural network demonstrates an improvement in average fidelity as compared to other reconstruction methods~\cite{Lohani2020}. We remark that our scheme aims to simultaneously reconstruct both the quantum state and the noise using incomplete measurements, by leveraging the low-rank property of the state and sparsity of the noise, while ensuring that the reconstructed quantum state is physical. If one focuses solely on the quality of state reconstruction, a detailed comparison between our scheme and the compressed sensing scheme can be found in~\ref{App:CS}.

In principle, the nonzero entries of the sparse noise vector $\bv$ may be arbitrarily large~\cite{8292873}. Taking into account different scenarios, our protocol can be adapted to handle noise through further post-processing if necessary, such as by imposing a bound on the noise vector or simply rescaling it. We expect that our approach can be combined with other techniques, leading to a more precise characterization method capable of effectively handling noise.

\section{Conclusion and outlook}\label{Sec:Summary}
In this work, we have proposed a scheme of corrupted sensing quantum state tomography and investigated its application under the assumption of \emph{low-rank} density matrix and \emph{sparse} statistical noise. Specifically, extensive numerical simulations were employed to demonstrate the simultaneous tomography performance of five-qubit random pure states under corrupted noise characterized by sparse Gaussian and Poisson noise, respectively. Our findings showed that by utilizing a limited number of quantum states and an incomplete set of measurement operators, the fidelity of the reconstructed state can reach as high as ${F(\rho, \hat{\rho})\gtrsim 0.95}$, and the MSE of the reconstructed noise is found to be in the order of $10^{-3}$. Expectedly, the reconstruction performance can be further improved by increasing the number of input quantum states and measurement operators.

Furthermore, we examined the robustness of the protocol by exploring different levels of the corrupted noise as well as random states with higher ranks. On top of that, we assessed the protocol's efficacy of the input states with local depolarizing noise and for larger systems with an increasing number of qubits. The results showcased the robustness and practicality of the protocol in real-world scenarios, especially for low-rank quantum states and relatively sparse noise. These findings offer valuable insights into the noise diagnosis and characterization of quantum systems.

We would like to highlight several interesting potential directions to explore in future. Firstly, we stress that the success of the algorithm is contingent on that proper values of the free parameters $\tau_1, \tau_2$ can be found, and what are the optimal parameters is left as an open problem. Secondly, the simple {\sc cvx} tool was employed as the initial solution for the protocol, while future work can explore more advanced techniques. Lastly, it is also essential to consider more diverse and intricate types of noise beyond statistical noise.

On the theoretical side, it would be necessary to derive a tight lower bound for the number of measurements required for the successful reconstruction of the quantum state and corrupted noise simultaneously. Instead of using restricted isometry property (RIP)~\cite{liu2011universal} tools to assess random Pauli measurements as in compressed sensing, the use of \emph{extended matrix deviation inequality}~\cite{Chen.Liu2019} or \emph{generalized RIP}~\cite{li2013compressed,8292873} may be explored to build the \emph{extended measurement matrix}. Furthermore, a concrete analysis of the error bound is warranted to provide insights into the algorithm's precision and limitations.

\section{Acknowledgments}
We are grateful to Yulong Liu and Yinfei Li for helpful discussions.
This work was supported by the National Natural Science Foundation of China (Grants No.~92265115 and No.~12175014) and the National Key R\&D Program of China (Grant No.~2022YFA1404900).

\section{Data availability statement}
The core code of this work can be found in the publicly available repository at \href{https://github.com/cQST-24/CorpSenQST}{https://github.com/cQST-24/CorpSenQST}. All data that support the findings of this study are included within the article (and any supplementary files).

%

\appendix

\section{The constrained setting}\label{App:Constrained}
\begin{figure}[t]
\centering
    \includegraphics[width=.5\columnwidth]{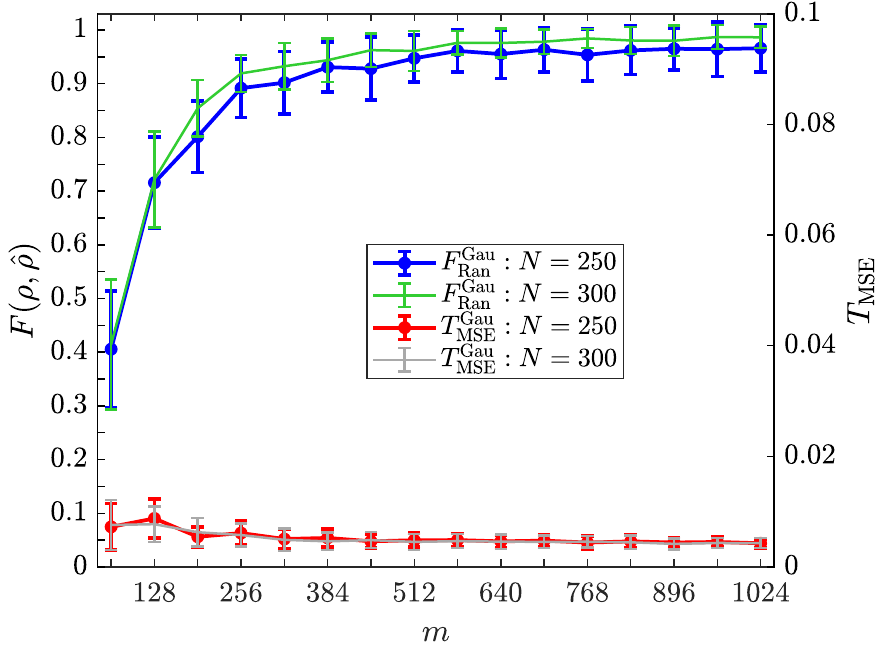}
\caption{Fidelity $F(\rho,\hat{\rho})$ and MSE $T_{\text{MSE}}$ as functions of the number of sampled Pauli operators $m$ over $120$ runs with $n=5$ qubits, when prior knowledge of the corrupted noise and the noise strength $\delta$ of $\bz$ are available. The input pure states are chosen randomly and the standard deviation of the corrupted Gaussian noise is set to ${\sigma =1}$. The sparsity level is set to ${s=\lfloor 0.04m \rfloor}$. Other parameters are chosen as  ${\beta=2.5}$ and ${\delta=0.1}$.}
\label{Sfig1}
\end{figure}
When prior knowledge of the corrupted noise $\bv$ (in terms of the $\ell_1$-norm) and the noise strength $\delta$ (in terms of the $\ell_2$-norm) of the unstructured noise $\bz$ are available, we can consider the following constrained recovery estimator
\begin{equation}\label{eq:Ala1}
    \minover{\tilde{\rho}\succeq0, \tilde{\bv}}\  \|\tilde{\rho}\|_{\tr}, \quad
    \subto \ \|\tilde{\bv}\|_1\le\beta\cdot\|\bv\|_1,\ \|\by-\cM(\tilde{\rho})-\tilde{\bv}\|_2\le \delta\,.\nonumber
\end{equation}
In addition, the reconstructed quantum state needs to be renormalized to ensure unit trace.

Here we provide an explanation for the constraint condition of the corrupted noise. When the $\ell_1$-norm of the corrupted noise is given as prior information, one can give a constraint condition ${\|\tilde{\bv}\|_1\le\|\bv\|_1}$. However, this condition is relatively strict, potentially resulting in a  small solution space for the optimization problem.  Therefore, we consider a looser constraint, namely ${\|\tilde{\bv}\|_1\le\beta\cdot\|\bv\|_1}$, where ${\beta=2.5}$ was selected based on numerical investigation. If more accurate noise reconstruction is desired, $\beta$ should be set to a smaller value (preferably ${\beta=1}$). Here, our primary focus is on the quality of state reconstruction.

We consider generating a Gaussian noise vector $\bz$ and scale $\bz$ such that ${\|\bz\|_2=\delta}$ for the data collection. Unstructured noise is commonly regarded as statistical noise due to the finite number of samples or disturbance introduced during the measurement process. Therefore, we express the constraint condition as ${\|\by-\cM(\tilde{\rho})-\tilde{\bv}\|_2\le \delta}$ with the aim of considering these noises by restricting the distance between the observed and the theoretical data.

In figure~\ref{Sfig1}, with ${N=250}$ state copies, the fidelity of the reconstructed state is ${F(\rho, \hat{\rho})\gtrsim 0.95}$ for ${m\gtrsim 576}$ measurement operators, and the corresponding MSEs are on the order of ${4.5\times 10^{-3}}$. Upon increasing the number of copies to $N=300$, a fidelity of $0.96$ is achieved with ${m=448}$ operators, and ${F(\rho, \hat{\rho})\approx0.986}$ with the complete set of measurement operators. The MSE is comparable to the case of ${N=250}$.
By the same token, one can also consider the scenario when prior knowledge of the quantum state and noise strength of the unstructured noise are known to minimize the $\ell_1$-norm of the corrupted noise.

\section{The penalized setting}\label{App:Penalized}
\begin{figure}[t]
\centering
\includegraphics[width=.5\columnwidth]{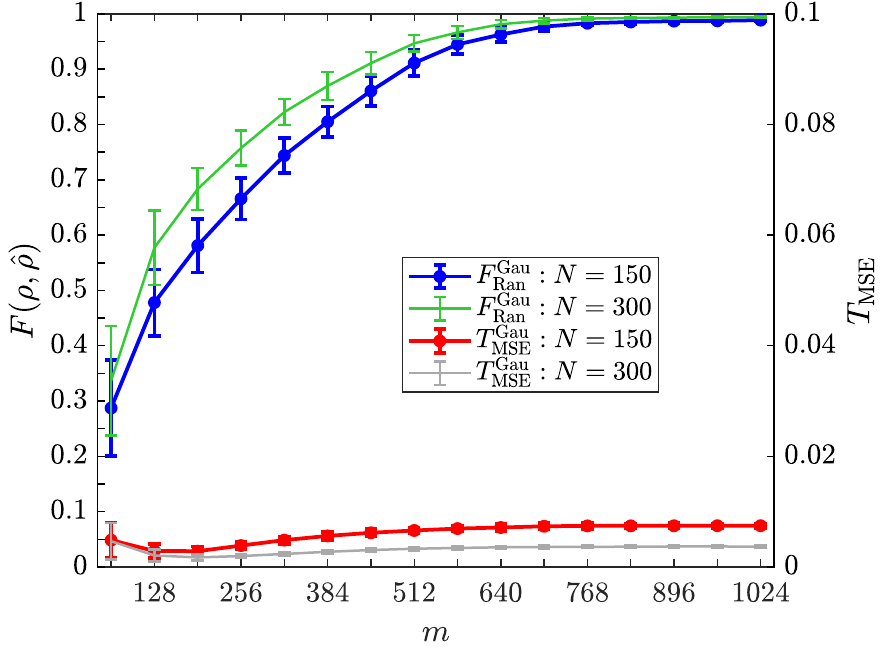}
\caption{Fidelity $F(\rho,\hat{\rho})$ and MSE $T_{\text{MSE}}$ as functions of the number of sampled Pauli operators $m$ over $120$ runs with ${n=5}$ qubits, when prior knowledge of the noise strength $\delta$ of $\bz$ is available. The input pure states are chosen randomly and the standard deviation of the corrupted Gaussian noise is set to ${\sigma=1}$. The sparsity level is taken as ${s=\lfloor 0.04m \rfloor}$. The regularization parameters are chosen as ${\lambda_1 = 0.001m, \lambda_2 = 0.014 + 0.002\cdot\frac{m-64}{m}}$, and ${\delta=0.1}$.}
\label{Sfig2}
\end{figure}

When only the noise strength $\delta$ of the unstructured noise $\bz$ is known, it is convenient to use the following penalized recovery estimator
\begin{equation}\label{eq:Alb1}
\minover{\tilde{\rho}\succeq0,\tilde{\bv}} \ \lambda_1\cdot\|\tilde{\rho}\|_{\tr}+\lambda_2\cdot\|\tilde{\bv}\|_1,\quad
    \subto\ \|\by-\cM(\tilde{\rho})-\tilde{\bv}\|_2\le \delta, \ \lambda_1, \lambda_2 > 0\,.\nonumber
\end{equation}
Additionally, the reconstructed quantum state should be renormalized.

The results are shown in figure~\ref{Sfig2}. To a certain extent, there is also a \emph{trade-off} between the accuracy of the state reconstruction and precision of the structured noise recovery, which can be evaluated by choosing the free parameters $\lambda_1$ and $\lambda_2$.
The findings reveal that for the case of ${N=150}$, when the fidelity reaches  ${\sim 0.96}$, the
number of measurement operators to be used is ${m\sim640}$. And a fidelity of ${F(\rho, \hat{\rho})\approx0.99}$ can be reached by increasing the state copies to ${N=300}$. Meanwhile, all MSEs are in the order of $10^{-3}$.

\section{$W$ state with corruption}\label{App:Wstate}
\begin{figure}[t]
\centering
    \includegraphics[width=.5\columnwidth]{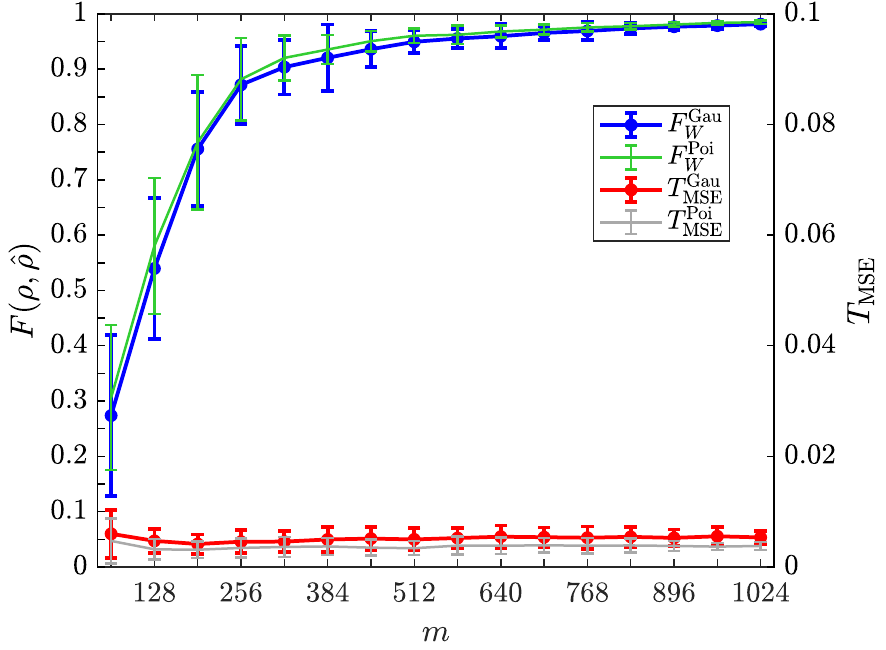}
\caption{Fidelity $F(\rho,\hat{\rho})$ and MSE $T_{\text{MSE}}$ as functions of the number of sampled Pauli operators $m$ for five-qubit $W$ state over $120$ runs. The number of copies used for each measurement is specified as ${N=100}$. The regularization parameters are chosen as ${\tau_1=0.011m, \tau_2=0.16}$. Standard deviation of the sparse Gaussian noise and parameter of the sparse Poisson noise are set to ${\sigma = \lambda =1}$. The sparsity level is taken as ${s=\lfloor 0.04m \rfloor}$.}
\label{Wstatefig}
\end{figure}
A similar simulation experiment is performed to test the $W$ state for a five-qubit system with corruption. An $n$-qubit $W$ state is written as
\begin{equation}\label{Wstate}
  \ket{W_n}=\frac{1}{\sqrt{n}}(\ket{10\cdots 0}+\ket{01\cdots 0}+\ket{00\cdots 1})\,.
\end{equation}
As can be seen in figure~\ref{Wstatefig}, variations of the fidelity and MSE are similar to those of the random pure states. Using the same parameter settings with sparse Gaussian noise and ${N = 100}$ state copies, the number of measurements required to achieve a fidelity of $0.95$ is about ${m=576}$, with the corresponding MSE ${T_{\text{MSE}}\approx 5.2\times 10^{-3}}$. Moreover, the fidelity can reach as high as ${F(\rho,\hat{\rho})\approx 0.98}$ when ${m=1024}$. In addition, it is worth mentioning that the parameter selection here may not be optimal, as the parameters can be adjusted depending on the scenarios being tested.

\section{Convergence with respect to $N$}\label{App:N}
\begin{figure}[t]
\centering
   \includegraphics[width=.5\columnwidth]{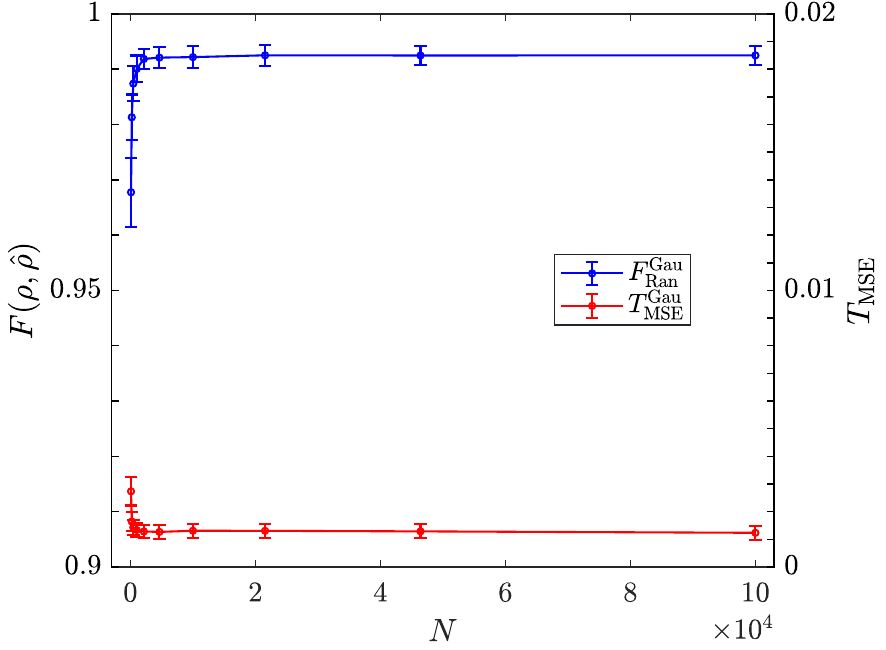}
    \caption{Fidelity $F(\rho,\hat{\rho})$ and MSE $T_{\text{MSE}}$ as functions of the state copy number $N$ over $120$ runs with ${n=5}$ qubits. The number of Pauli operators is fixed as ${m=512}$. Standard deviation of the sparse Gaussian noise is set to ${\sigma=1}$. The regularization parameters are chosen as ${\tau_1=0.011m, \tau_2=0.16}$. The sparsity level is taken as ${s=\lfloor 0.04m \rfloor}$.}
    \label{fig_N}
\end{figure}
In figure~\ref{fig_N}, the fidelity $F(\rho,\hat{\rho})$ and MSE $T_{\text{MSE}}$ nearly converge when $\sim 5\times 10^3$ state copies are used for estimating the expectation value $\Tr(P_k\rho)$.
In detail, the fidelity $F(\rho,\hat{\rho})$ increases from approximately $0.9677$ to $0.992$ as $N$ changes from $100$ to $4642$, while the corresponding MSE $T_{\text{MSE}}$ decreases from around $0.0027$ to $0.0013$.

\section{Recovery noise error}\label{App:noiseError}
\begin{figure}[t]
\centering
 \includegraphics[width=.5\columnwidth]{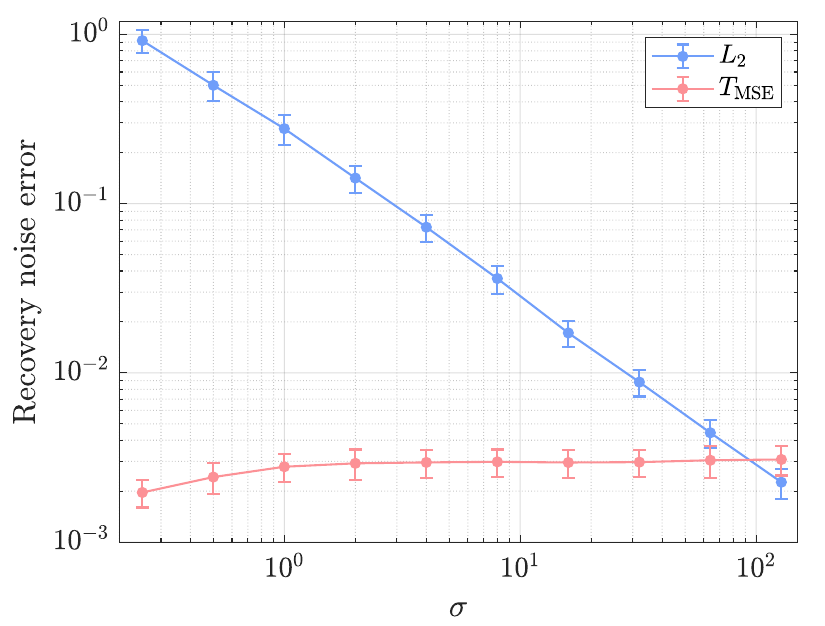}
    \caption{Relative $\ell_2$-norm error $L_2$ and MSE ${T_{\text{MSE}}}$ as functions of the standard deviation of the sparse Gaussian noise $\sigma$ over $120$ runs with ${n=5}$ qubits. The number of copies of the input random pure states used for each measurement is specified as ${N=100}$ and the number of Pauli operators is fixed as ${m=512}$. The regularization parameters are chosen as ${\tau_1=0.011m, \tau_2=0.16}$. The sparsity level is taken as ${s=\lfloor 0.04m \rfloor}$.}
\label{noiseErrorfig}
\end{figure}
We define the relative $\ell_2$-norm error $L_2$ as
\begin{equation}
    L_2=\frac{\|\hat{\bv}-\bv\|_2}{\|\bv\|_2}\,.
\end{equation}
Figure~\ref{noiseErrorfig} shows the recovery noise error measured by $L_2$ and ${T_{\text{MSE}}}$ as functions of the standard deviation of the sparse Gaussian noise. The standard deviation $\sigma$ is varied within the range ${[2^{-2}, 2^{7}]}$, increasing by the powers of $2$. The relative $l_2$-norm error decreases from ${L_2\approx 0.92}$ to ${L_2\approx2.3 \times 10^{-3}}$.

\section{Larger systems with more qubits}\label{App:7qubit}
\begin{figure}[t]
\centering
   \includegraphics[width=.5\columnwidth]{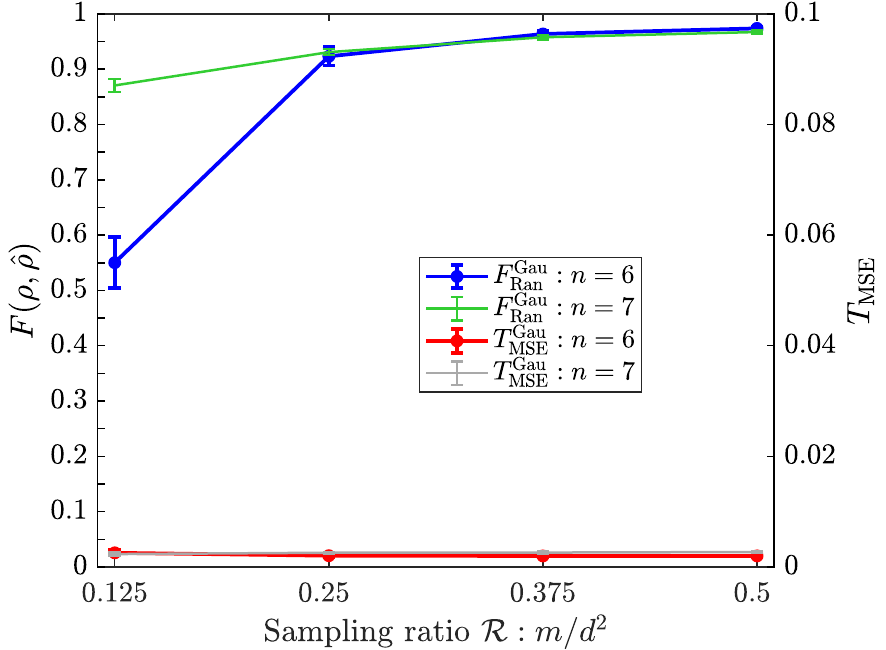}
    \caption{Fidelity $F(\rho,\hat{\rho})$ and MSE $T_{\text{MSE}}$ as functions of the sampling ratio $\mathcal{R}$ over $20$ runs with ${n=6,7}$ qubits. The number of copies of the input random pure states used for each measurement is specified as ${N=100}$. Standard deviation of the sparse Gaussian noise is set to ${\sigma=1}$. The regularization parameters are chosen as ${\tau_1=0.0055m, \tau_2=0.16}$. The sparsity level is taken as ${s=\lfloor 0.04m \rfloor}$.}
    \label{fig_q}
\end{figure}
To demonstrate the potential applications of our protocol, the number of qubits is increased in figure~\ref{fig_q}. When ${n = 6}$, the fidelity at ${\mathcal{R}=0.375}$ is ${F(\rho,\hat{\rho})\approx 0.96}$, where ${\mathcal{R}=m/d^2}$; for ${n = 7}$, ${F(\rho,\hat{\rho})\approx 0.958}$. The corresponding MSE are $0.002$ and $0.0026$, respectively. As stated in section~\ref{subsec4.1}, a fidelity of $0.95$ can be achieved with a sampling rate of $37.5\%$.

\section{Comparison with compressed sensing}\label{App:CS} 
We now focus on the quality of state reconstruction. Consider the output of equation~(\ref{eq:corSens}), while we employ the matrix Lasso estimator from the compressed sensing scheme~\cite{Flammia.etal2012} as given below, to reconstruct the state only.
\begin{equation}\label{eq:cs}
    \hat{\rho} = \mathop{\arg\min}_{\tilde{\rho} \succeq 0}  \frac{1}{2}\|\by - \cM(\tilde{\rho})\|_2^2 + \mu \cdot \|\tilde{\rho}\|_{\tr}, \  \mu > 0\,.
\end{equation}
 Here we generate structured noise $\bv$ with $s$ non-zero elements following a standard Gaussian distribution, and bound the noise such that ${\|\bv\|_2=\delta_v}$, where ${\delta_v=\delta_0\sqrt{s}}$ and ${s=\lfloor \eta m \rfloor}$. This bound on the noise level is used to ensure the validity of the compressed sensing estimator. When ${\eta=1}$, the noise $\bv$ is completely unstructured. The regularization parameter $\mu$ is chosen as ${\mu = 0.011m}$. Additionally, the reconstructed quantum state is renormalized to ensure unit
trace. For comparison, under the same noise settings, we also present the results of our scheme as in equation~(\ref{eq:normgiven}), where the regularization parameters $\tau_1, \tau_2$ are set to be the same as in section~\ref{subsec4.1}. The performance of our scheme for noise reconstruction can be found in section~\ref{Sec:Applications} and~\ref{App:noiseError}. Here, we only concentrate on state reconstruction.

\begin{figure}[t]
\centering
    \includegraphics[width=.6\columnwidth]{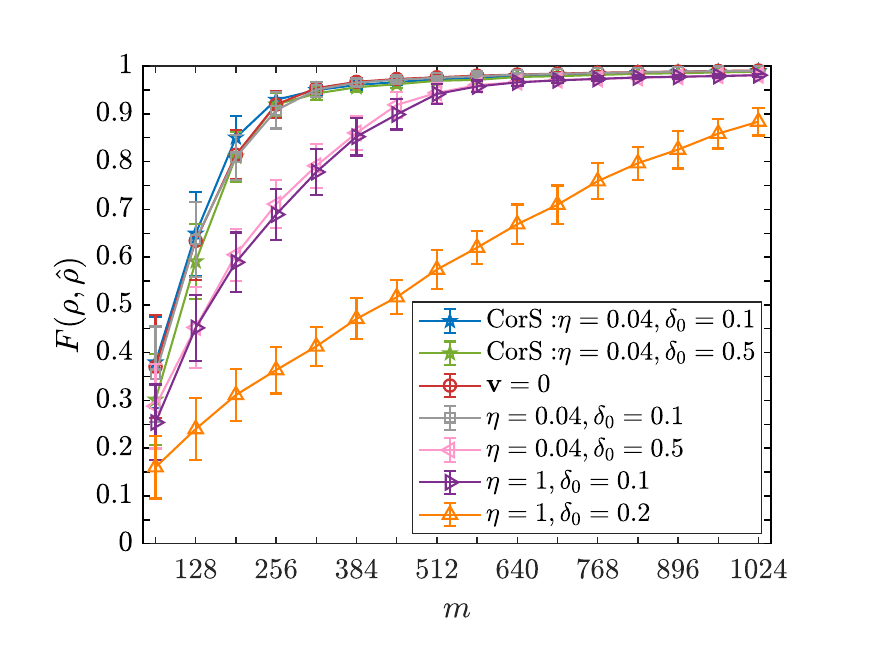}
\caption{Fidelity $F(\rho,\hat{\rho})$ as a function of the number of sampled Pauli operators $m$ over $120$ runs with ${n=5}$ qubits using different estimators. The number of
copies of the input random pure states used for each measurement is specified as ${N = 100}$. The blue and green curves show the fidelity of the corrupted sensing scheme, while the other curves represent the results obtained using the matrix Lasso estimator.}
\label{figG1}
\end{figure}

The fidelity as a function of the number of sampled Pauli operators is shown in figure~\ref{figG1}. For sparse noise $\bv$ with ${\eta=0.04}$, it can be observed that as the noise level $\delta_0$ increases, the corrupted sensing scheme (blue and green curves) exhibits robustness as compared to the matrix Lasso scheme (gray and pink curves). In the matrix Lasso scheme, the best reconstruction occurs when the noise ${\bv=0}$ (red curve), followed by the case of structured noise (${ \eta = 0.04}$), and the worst performance is observed with unstructured noise (${\eta=1}$, represented by the purple and orange curves). 
It is expected that the state reconstruction improves as the noise level decreases in the compressed sensing scheme. On the contrary, in our scheme, the noise $\bv$, of which we aim to reconstruct simultaneously, is not a burden.
%

\bibliographystyle{iopart-num}
\section*{References}

\providecommand{\newblock}{}

\end{document}